\documentclass[aps,prl, twocolumn,superscriptaddress,amssymb,amsmath,nofootinbib]{revtex4-1}

\usepackage{color}
\usepackage{epsfig}
\usepackage{graphicx}
\usepackage{float}
\usepackage{epstopdf}
\usepackage{hyperref} 
\usepackage[usenames,dvipsnames]{xcolor}
\usepackage{multirow}
\usepackage[normalem]{ulem}
\usepackage[d]{esvect}

\begin{document}

\begin{flushright}
	EFI-21-7
\end{flushright}
\title{$\nu$ scalar in the early Universe and $(g-2)_{\mu}$ }

\author{Jia Liu}
\email{jialiu@pku.edu.cn}
\affiliation{School of Physics and State Key Laboratory of Nuclear Physics and Technology, Peking University, Beijing 100871, China}
\affiliation{Center for High Energy Physics, Peking University, Beijing 100871, China}

\author{Navin McGinnis}
\email{nmcginnis@triumf.ca}
\affiliation{TRIUMF, 4004 Westbrook Mall, Vancouver, BC, Canada, V6T 2A3}

\author{Carlos E.M. Wagner}
\email{cwagner@uchicago.edu}
\affiliation{High Energy Physics Division, Argonne National Laboratory, Argonne, IL 60439, USA}
\affiliation{\mbox{Physics Department and Enrico Fermi Institute, University of Chicago, Chicago, IL 60637}}
\affiliation{\mbox{Kavli Institute for Cosmological Physics, University of Chicago, Chicago, IL 60637, USA}}

\author{Xiao-Ping Wang}
\email{Corresponding author. hcwangxiaoping@buaa.edu.cn}
\affiliation{School of Physics, Beihang University, Beijing 100083, China}
\affiliation{Beijing Key Laboratory of Advanced Nuclear Materials and Physics, Beihang University, Beijing 100191, China}


\date{\today}

\begin{abstract} 
We investigate a concrete scenario of a light scalar with a mass around 1~MeV which can be connected to the origin of neutrino masses and simultaneously survive current bounds on relativistic degrees of freedom in the early universe. A feeble coupling to the Standard Model neutrinos can relax stringent bounds on the decays to photons inferred from the measured value of $N_{\rm eff}$. Interestingly, we find that such a scalar whose diphoton coupling is generated by a tree-level coupling to the muon of similar strength as that of the Standard Model Higgs boson can simultaneously explain the longstanding discrepancy in the measured value of the muon magnetic moment. We present a possible ultraviolet (UV) completion of this scenario providing a link between new physics in the early universe and the generation of neutrino masses. 
\end{abstract}

\pacs{}

\keywords{}

\maketitle

\noindent {\textbf{Introduction}}--
The theoretical and experimental establishment of the Standard Model (SM) of particle physics~\cite{ParticleDataGroup:2020ssz} and general relativity ushered in
the so-called $\Lambda {\rm CDM}$ model,
which provides an excellent description of nucleonsynthesis and cosmic microwave background data~\cite{Trodden:2004st}. The existence of particles beyond the SM can modify the expansion rate of the universe, impacting both observables~\cite{Olive:1998vj, Hannestad:2003ye}.
Recently, several detailed analyses have explored the constraints on new light particles in the early universe~\cite{Millea:2015qra, Sabti:2019mhn, Depta:2020wmr, Carenza:2020zil, Giovanetti:2021izc, Batell:2021ofv} where big bang nucleosynthesis (BBN) and Planck data imply strong constraints on (pseudo) scalar particles with an MeV-scale mass and coupling to photons, mainly due to the corresponding contributions to $N_{\rm eff}$. However, as emphasized, for instance, in Refs.~\cite{Sabti:2019mhn,Depta:2020wmr,Huang:2021dba}, these bounds are not robust and could be modified either due to the details regarding the generation of the diphoton coupling or those of the possible modifications to the neutrino or, in general, light particle sector. 

New light  scalars, as well as new, sterile neutrinos may also be connected to the mechanism generating neutrino masses, which remains an open question in particle physics. Additionally, the current observation of a deviation of the anomalous magnetic moment~\cite{Muong-2:2021ojo} from its theoretically predicted value in the SM~\cite{Davier:2010nc, Aoyama:2012wk, Aoyama:2019ryr, Czarnecki:2002nt,Gnendiger:2013pva,Davier:2017zfy,Keshavarzi:2018mgv,Colangelo:2018mtw,Hoferichter:2019gzf,Davier:2019can,Keshavarzi:2019abf,Kurz:2014wya,Melnikov:2003xd,Masjuan:2017tvw,Colangelo:2017fiz,Hoferichter:2018kwz,Gerardin:2019vio,Bijnens:2019ghy,Colangelo:2019uex,Blum:2019ugy,Colangelo:2014qya} (see a recent review \cite{Aoyama:2020ynm}) demands
an explanation and could be connected to the presence of new scalars coupled to 
the muon~\cite{Kinoshita:1990aj, Zhou:2001ew, Barger:2010aj, Tucker-Smith:2010wdq, Chen:2015vqy, Batell:2016ove, Liu:2018xkx, Caputo:2021rux}.

Motivated by these observations, in this paper we present a low-energy model of a real scalar particle with mass $\mathcal{O}(1)$ MeV which, at tree level, couples to the muon and SM neutrinos, and whose UV completion provides the source of neutrino masses. Interestingly, we find that a scalar with mass in this range 
can simultaneously generate the necessary correction to explain the recently measured value of the muon magnetic moment~\cite{Muong-2:2021ojo}, 
\begin{equation}
\Delta a_{\mu}=(2.51\pm 0.59)\times 10^{-9},
\label{eq:amurange}
\end{equation}
and, for an appropriate coupling to neutrinos, satisfy all cosmological, astrophysical and laboratory constraints. 
\\

\noindent {\textbf{Effective Model}}--
We concentrate on the properties of a real singlet scalar which at low energies couples exclusively to the SM neutrinos and the muon
\begin{equation}
\label{eq:Leff}
\mathcal{L}_{\rm eff} \supset  - g_{\mu}\phi\bar{\mu}\mu -\left(\frac{g_{\nu_a}}{2}\phi  \ \nu_{a}\cdot\nu_{a} + H.c.\right),
\end{equation}
where, $\nu_a$ are the active neutrino flavors and, we have used Dirac notation for the first term versus Weyl notation for the second.
As anticipated, the $\phi$ coupling to neutrinos provides the source of the neutrino Majorana masses.
The muon mass, instead, comes mostly from Higgs induced, $\phi$ independent terms within the UV completion of the model.

The coupling to the muon will induce the coupling to photons at one-loop~\cite{Delbourgo:2000nq}
\begin{equation}
\mathcal{L}_{\rm eff} \supset- \frac{g_{\gamma\gamma}}{4}\phi F_{\mu\nu}F^{\mu\nu},
\end{equation}
with $g_{\gamma\gamma} \approx -(2\alpha g_{\mu})/(3\pi m_{\mu})$ for $m_{\phi}\ll m_{\mu}$~\cite{Delbourgo:2000nq}.

The scalar decay widths to photons and neutrinos are then given by
\begin{equation}
\Gamma_{\gamma\gamma}=\frac{g_{\gamma\gamma}^{2}m_{\phi}^{3}}{64\pi},
\quad \Gamma_{\nu\nu}=\sum_a \frac{g_{\nu_a}^{2}m_{\phi}}{32\pi},
\end{equation}
respectively. It is clear that the same parameters, $g_{\mu}$ and $m_{\phi}$, which fix the coupling to photons,  simultaneously determine the contribution of $\phi$ to the muon anomalous magnetic moment 
which at one-loop is given by
\begin{equation}
\Delta a_{\mu}=\frac{g_{\mu}^{2}}{8\pi^2}\int_{0}^{1} dz\frac{(1-z)^{2}(1+z)}{(1-z)^{2}+z(m_{\phi}/m_{\mu})^{2}}.
\end{equation}
Thus, the experimentally determined range for $\Delta a_{\mu}$, Eq.~(\ref{eq:amurange}), leads directly to a prediction for $g_{\gamma\gamma}$. For $m_{\phi}\lesssim 1$ MeV, this is achieved at the 2-$\sigma$ level if $g_{\mu}\simeq m_{\mu}/v$, similar to the Higgs coupling to muons in the SM~\cite{Liu:2018xkx}.

\begin{figure}[thb]
	\centering
	\includegraphics[width=0.48 \textwidth]{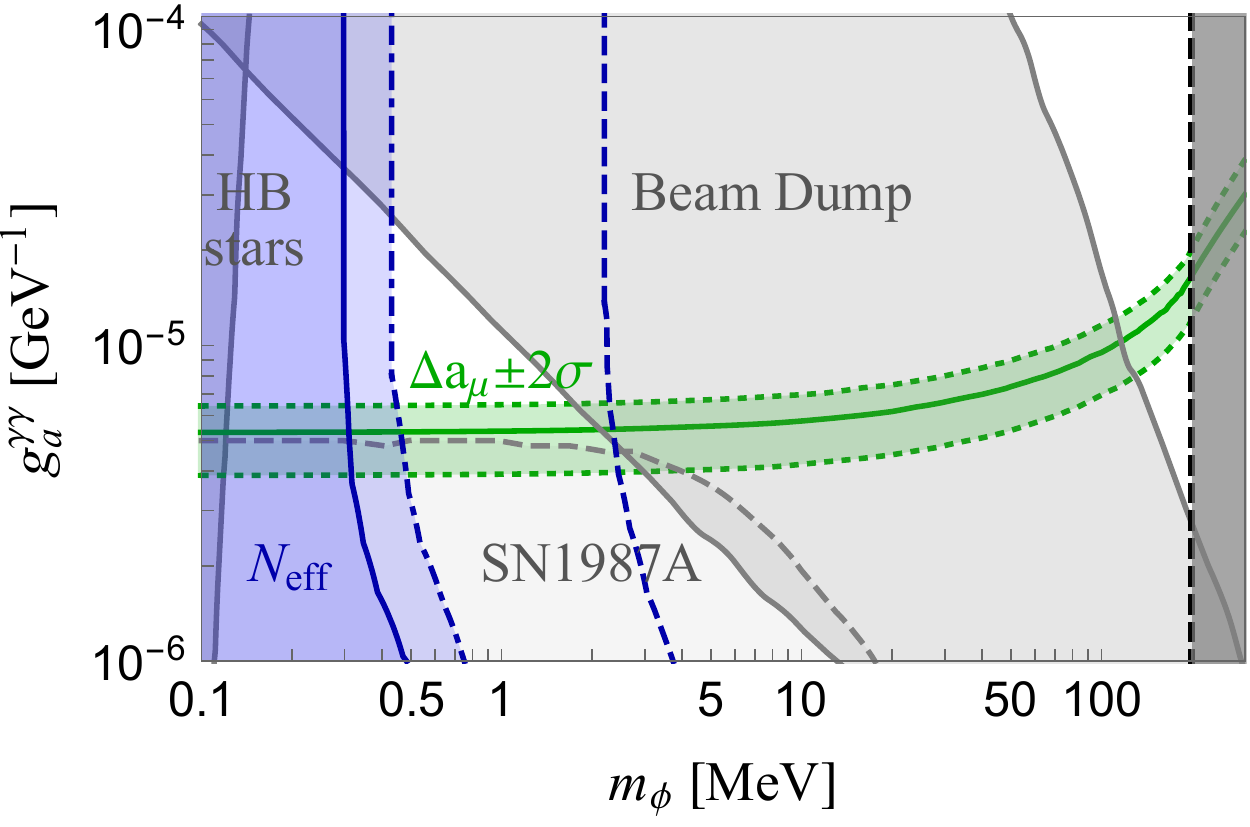}
	\caption{Allowed region of $g_{\gamma\gamma}$ [GeV$^{-1}$] vs $m_{\phi}$ [MeV] with constraints taken from~\cite{Depta:2020wmr}. The green shaded region shows the values of $g_{\gamma\gamma}$ consistent with $\Delta a_{\mu}$ within 2$\sigma$. The dashed line shows the bound based on the standard cosmology. Dot-dashed and solid lines show the effects of extra relativistic d.o.f. and a neutrino chemical potential, respectively.}
	\label{fig:gm2_neff}
\end{figure}

In Fig.~\ref{fig:gm2_neff}, we show the allowed of $g_{\gamma\gamma}$ for a scalar with mass $\sim \mathcal{O}(0.1-100)$ MeV where constraints are presented from both laboratory and astrophysical searches, as presented in~\cite{Depta:2020wmr}. 
Let us emphasize that,
although Ref.~\cite{Depta:2020wmr} concentrates in the case of pseudoscalars, as these constraints are derived with respect to the lifetime and mass of these particles, they will be equally stringent for scalar particles. 
While the constraints from beam dumps and HB stars (shown in the gray shaded regions with solid-line boundaries) 
should be considered robust, we note that the bound from SN1987A (gray shaded region with a dashed-line boundary) has recently received critical skepticism in the literature and could very well be much weaker~\cite{Lee:2018lcj,Bar:2019ifz,Lucente:2020whw}. On the other hand, it was emphasized recently that this region of parameters could be excluded from supernova (SN) bounds resulting from muon-specific SN models~\cite{Caputo:2021rux}. We have not included their explosion energy bounds as the addition of the neutrino coupling will modify the trajectory of $\phi$ within the neutrino sphere itself and the muonic density profile in current SN models do not include the possible new physics effects.

The green shaded region in Fig.~\ref{fig:gm2_neff} shows the predicted values of $g_{\gamma\gamma}$ when $g_{\mu}$ is fixed to satisfy $\Delta a_{\mu}$ within 2$\sigma$. 
The bounds from $N_{\rm eff}$ constrain masses up to $m_{\phi}\simeq 2$~MeV assuming the standard cosmological evolution. The blue shaded regions show the weaker constraints associated with a modification of the cosmological evolution by the addition of extrarelativistic degrees of freedom (d.o.f.) or a neutrino chemical potential. Taking into account the possible modification of $N_{\rm eff}$ leaves an open region for $m_\phi$ in $1$--$2$ MeV, which has been referred to as
\textit{the cosmological triangle}.
In contrast to~\cite{Depta:2020wmr}, the scenario we consider leads to a similar variation of $N_{\rm eff}$ without any additional new physical degree of freedom by considering additional decays of $\phi$ to SM neutrinos.

Finally, it should be emphasized that there exists a window of masses $m_{\phi}\simeq\mathcal{O}(100)$ MeV which is currently not subject to either SN or $N_{\rm eff}$ constraints. Rather, this region is bounded from below by beam dump experiments, and above by $2m_{\mu}$ where the threshold for pair production of muons sets an upper bound where recent constraints from BABAR and Belle-II become relevant~\cite{BaBar:2020jma,Campajola:2021pdl}. 
\\

\noindent {\textbf{$N_{\rm eff}$ calculation}}--
In the standard neutrino cosmology picture, the effective number of neutrino species can be parameterized by the ratio between neutrino and photon densities after neutrino decoupling and electron-positron annihilation in the thermal background~\cite{ParticleDataGroup:2020ssz}
\begin{equation}
\frac{\rho_{\nu}}{\rho_{\gamma}}=\frac{7}{8}N_{\rm eff}\left(\frac{T_{\nu}}{T_{\gamma}}\right)^{4}_{\rm SM},
\label{eq:Neff_def}
\end{equation}
where the SM value, $(T_{\nu}/T_{\gamma})_{\rm SM}=(4/11)^{1/3}$, is determined by requiring that the neutrino and electromagnetic plasma separately conserve comoving entropy.

In our model, additional production mechanisms, e.g. $\mu^+ \mu^-  \leftrightarrow  \gamma \phi$,
$\gamma \ell^\pm \leftrightarrow \phi \ell^\pm$ and $\gamma\gamma, \nu\nu \leftrightarrow \phi$, where $\ell^\pm$ stand for the charged leptons, make $\phi$ fully thermalized before BBN. After neutrino decoupling, $\phi$ will inject entropy separately into the neutrino and electromagnetic plasmas, whose entropies themselves will evolve independently. $N_{\rm eff}$ may be determined by the requirement of conservation of co-moving entropy alone~\cite{Boehm:2013jpa}. Thus, rather than resort to a calculation of the scalar abundance from neutrino interactions, as in~\cite{Gelmini:1980re, Blinov:2019gcj, DeGouvea:2019wpf, Lyu:2020lps, Kelly:2020aks}, we may evalute the modified neutrino decoupling temperature $T_D$, which will in general be different than $T_{D}^{\rm SM} = 2.3$ MeV in the SM.

\begin{figure*}[tbh]
	\includegraphics[width=0.38\textwidth]{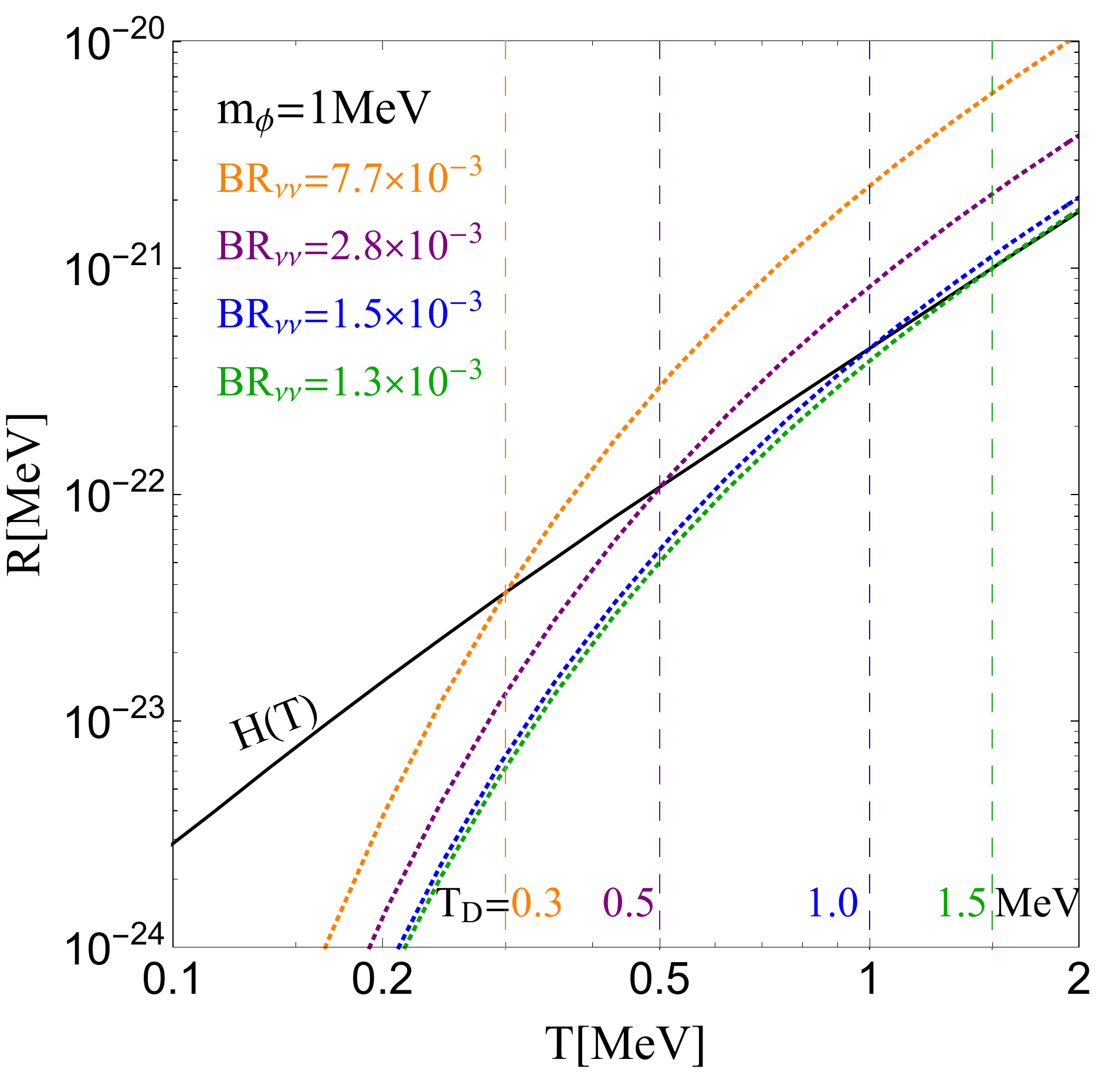}
	\includegraphics[width=0.38\textwidth]{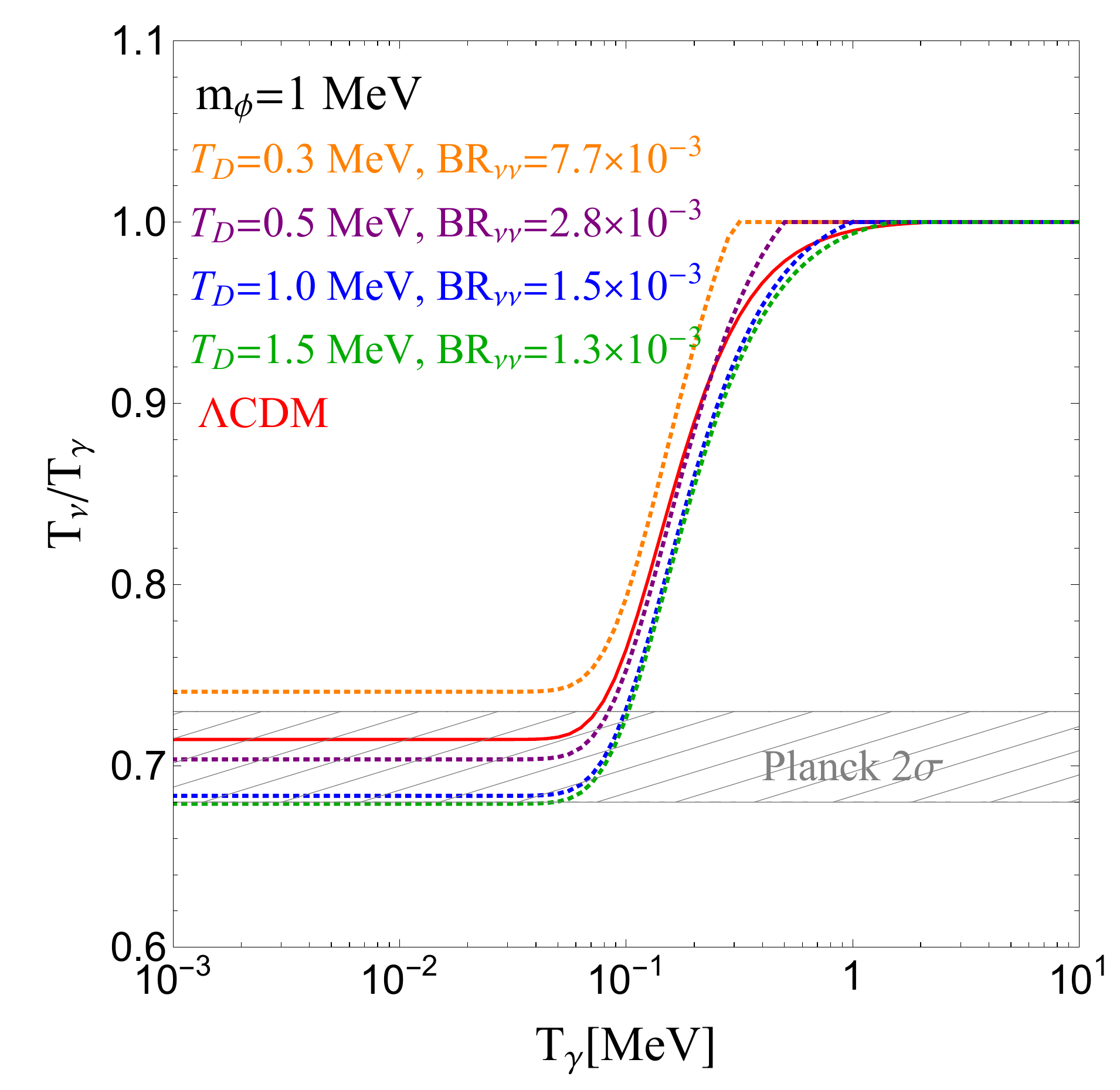}\\
	\includegraphics[width=0.38\textwidth]{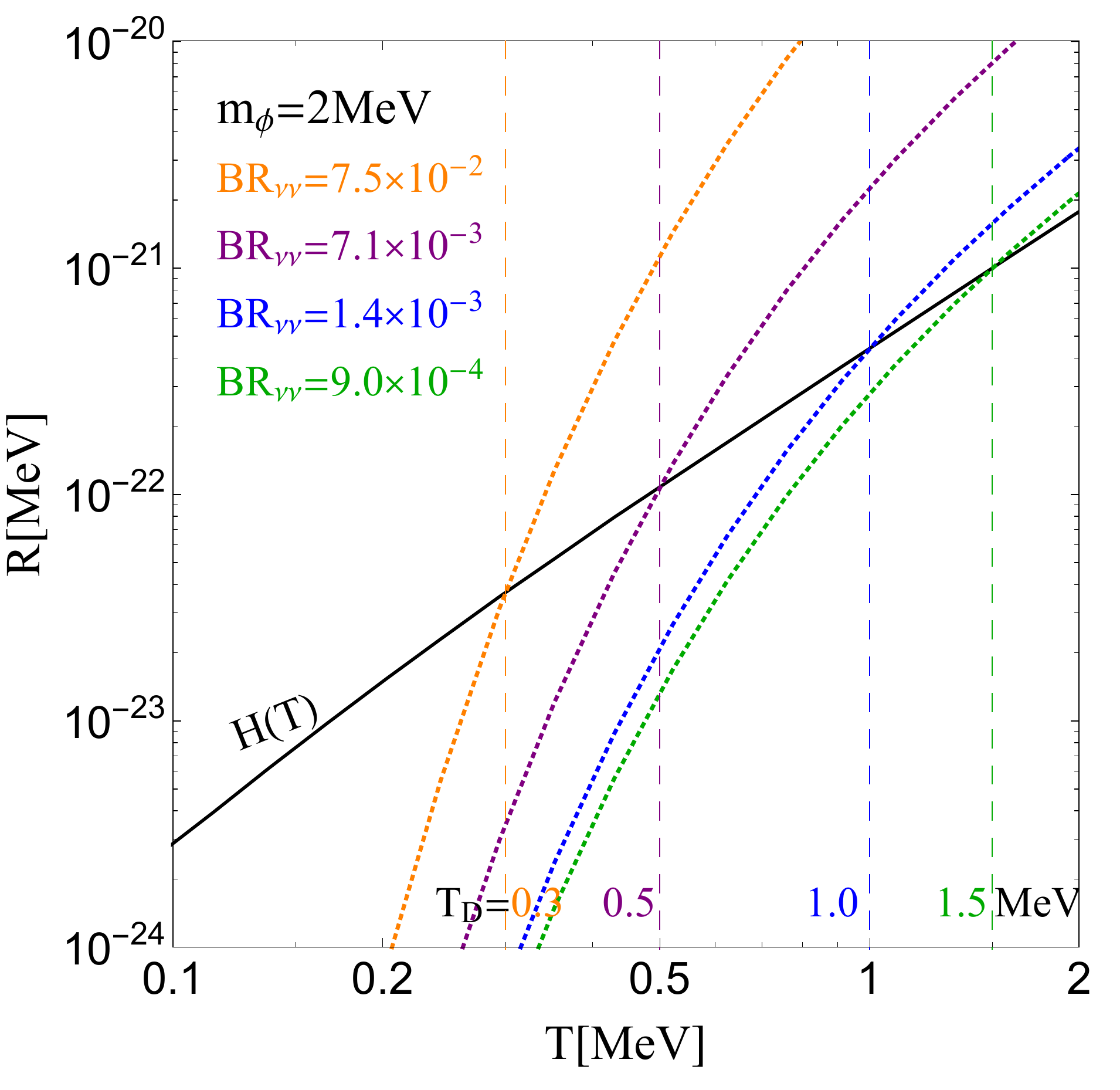}
	\includegraphics[width=0.38\textwidth]{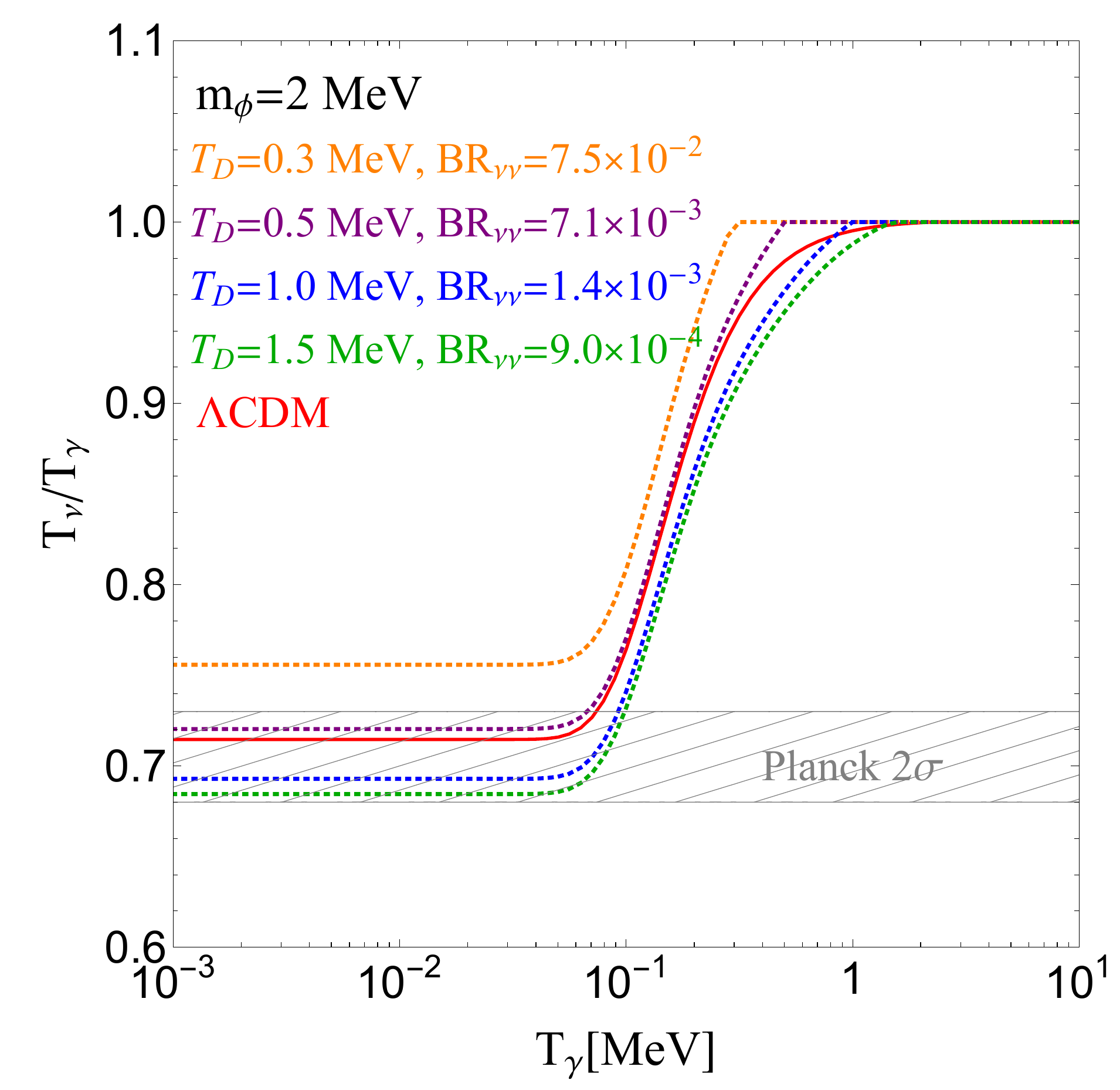}
	\caption{Left panel: comparison of the rate, $R$, of the process $\nu \nu \leftrightarrow \phi^{(*)} \leftrightarrow \gamma \gamma$ with that of Hubble, $H(T)$. Examples of $T_D$, determined via vertical dashed lines where $R\simeq H(T)$, are shown for different ${\rm BR}_{\nu \nu}$.
	Right panel:
	evolution of $T_\nu / T_\gamma$ as a function of the photon temperature
	for different ${\rm BR}_{\nu \nu}$. The red solid line and hatched gray region shows the evolution in ${\rm \Lambda{CDM}}$ and 2$\sigma$ Planck limits, respectively.
    The top (bottom) panels are for $m_\phi = 1$ (2) MeV respectively.}
	\label{fig:rate-Tratio}
\end{figure*}

The presence of new particles in thermal equilibrium with photons, electrons or neutrinos can alter $N_{\rm eff}$ through modifications of $T_{\nu}/T_{\gamma}$~\cite{Boehm:2013jpa}. For particles with mass $\mathcal{O}(1)$ MeV it follows that~\cite{Kolb:1986nf,Srednicki:1988ce,Boehm:2012gr, Boehm:2013jpa}
\begin{equation}
N_{\rm eff}=N_{\nu}\left(\frac{4}{11}\right)^{-4/3}\left(\frac{T_{\nu}}{T_{\gamma}}\right)^{4}.
\label{eq:neff}
\end{equation}
The requirement of conservation of comoving entropies after neutrino decoupling further determines the ratio $T_{\nu}/T_{\gamma}$ \cite{Boehm:2012gr}
\begin{equation}
\frac{T_{\nu}}{T_{\gamma}} =  \left(\frac{(g^{*}_{s})_{\nu}}{(g^{*}_{s})_{\gamma}} \Bigg|_{T_D} \frac{(g^{*}_{s})_{\gamma}}{(g^{*}_{s})_{\nu}}\right)^{1/3},
\end{equation}
where $(g^{*}_{s})_{\nu}$ and $(g^{*}_{s})_{\gamma}$ denote the effective number of relativistic degrees of freedom contributing to the neutrino and photon entropy densities, respectively. 

The neutrino decoupling temperature, $T_{D}$, is modified for scalar masses $\mathcal{O}(1)$ MeV via the presence of the s-channel resonant exchange $\nu \nu \leftrightarrow \gamma \gamma$ which delays neutrino decoupling, leading to lower $T_{D}$. The t-channel exchange $\gamma \nu \leftrightarrow \gamma \nu$ has a negligible contribution because of small couplings.
The Boltzmann equation for the process $\nu \nu \leftrightarrow \phi^{(*)}\leftrightarrow \gamma \gamma$ is given by
\begin{align}
	\frac{d n_\nu}{d t}+ 3H n_\nu = \langle \sigma v\rangle_{\rm res} \left( n_{\nu, {\rm eq}}^2- n_{\nu}^2\right),
\end{align}	
where the thermal averaged, resonant cross section is given by
\begin{align}
	\langle \sigma v\rangle_{\rm res} = \frac{16 \pi^2}{9 \xi(3)^2} 
	\Gamma_\phi  {\rm BR}_{\gamma\gamma} {\rm BR}_{\nu \nu} \frac{x^2}{T^3} K_1(x).
\end{align}
$\Gamma_{\phi}$ is the total width of $\phi$, the Riemann zeta function $\xi(3) = 1.202$, $x\equiv m_\phi/T$, $K_1(x)$ is the modified Bessel function of the second kind, ${\rm BR}_{\nu\nu} \equiv {\rm BR}(\phi \to \nu\nu)$ and ${\rm BR}_{\gamma\gamma}\equiv 1-{\rm BR}_{\nu\nu}$.
For $T \lesssim m_{\phi}$, we compare the exchange rate $R \equiv n_{\nu}^{\rm eq} \langle \sigma v \rangle_{\rm res}$ with the Hubble expansion rate and find the neutrino decoupling temperature

\begin{align}
	x_D \approx \log \left(  \frac{1.67 m_{\rm PL} \Gamma_\phi     }{ g_{*}^{1/2} m_\phi^2} {\rm BR}_{\gamma\gamma} {\rm BR}_{\nu \nu} \right) + \frac{7}{2} \log(x_D).
	\label{eq:TD_sol}
\end{align}

In the left panel of Fig.~\ref{fig:rate-Tratio}, we show the comparison of $R$ to the Hubble rate for values of the neutrino branching ratio ${\rm BR}_{\nu\nu}\simeq\mathcal{O}(10^{-2}) - \mathcal{O}(10^{-3})$. A larger branching ratio to neutrinos leads to lower values of $T_{D}$. In general, a lower $T_D$ than that in the SM will lead to entropy sharing between the photon and neutrino baths, in particular after $e^+ e^-$ annihilation if $T_{D}\lesssim  m_e$.  The resulting slower reheating of the photon bath from $e^+ e^-$ annihilation will cause the baryon number density to increase leading to strong constraints from BBN, $T_D \gtrsim 0.3$ MeV~\cite{Sabti:2019mhn}.
\begin{figure}[tbh]
	\includegraphics[scale=0.4]{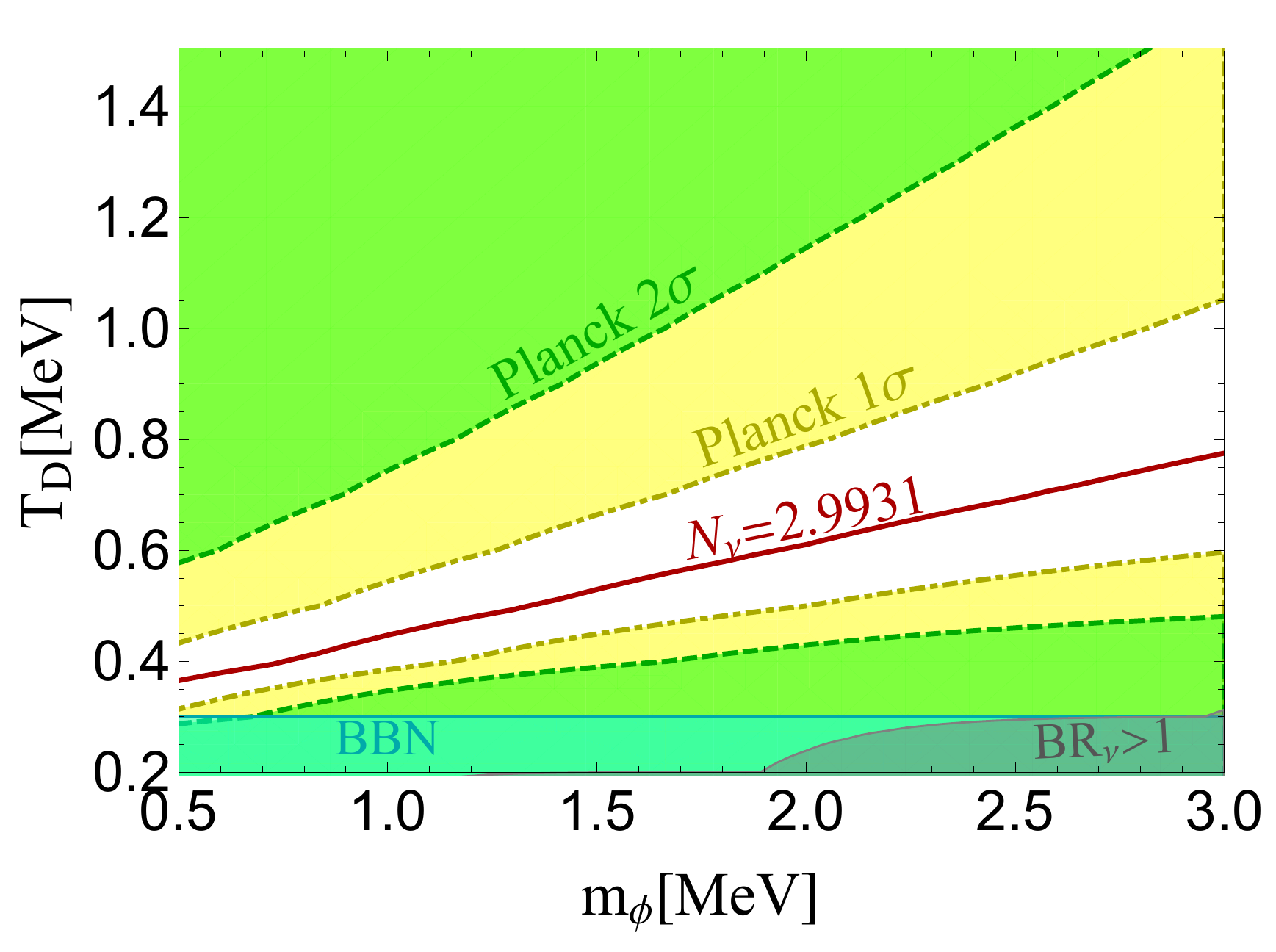}
	\caption{Limits on $T_D$ and $m_\phi$ from the Planck measurement of $N_{\rm eff}$ \cite{Planck:2018vyg}.
		The color shaded regions are excluded at 1$\sigma$ and 2$\sigma$ respectively.
		The lower bound $T_D > 0.3$ MeV \cite{Sabti:2019mhn} is plotted as shaded cyan.}
	\label{fig:TD2}
\end{figure}

At temperatures below $T_{D}$, the decay modes of $\phi$ will inject entropy into both the electromagnetic and neutrino sectors, while the annihilation of $e^+/e^-$ will further provide injection to the photon entropy. Requiring that the entropy in each sector is conserved, we can then evaluate the ratio $T_\nu / T_\gamma$ for a given branching ratio $\rm BR_{\nu\nu}$ and temperature, $T \leq T_D$,  
\small
\begin{align}
& \frac{T_\nu}{T_\gamma} \approx   \left( \frac{2+ \frac{7}{2}F(\frac{m_e}{T_\gamma})+\frac{7}{8} {\rm BR}_{\gamma\gamma} F(\frac{m_\phi}{T_\gamma})}{2+\frac{7}{2}F(\frac{m_e}{T_D})+\frac{7}{8} {\rm BR}_{\gamma\gamma} F(\frac{m_\phi}{T_D})} \cdot
	\frac{N_\nu + \frac{1}{2} {\rm BR}_{\nu\nu} F(\frac{m_\phi}{T_D}) }{N_\nu + \frac{1}{2} {\rm BR}_{\nu\nu} F(\frac{m_\phi}{T_\gamma})}
	\right)^{\frac{1}{3}} , \nonumber \\
&	F(y) = \frac{30}{7 \pi^4} \int_y^{\infty} dx \frac{(4x^2 -y^2) \sqrt{x^2-y^2}}{e^x \pm1},
\label{eq:TnuTgamma}
\end{align}
\normalsize
where the plus (minus) signs correspond to fermion (boson) particles.

In the right panel of Fig.~\ref{fig:rate-Tratio}, we show the evolution of $T_{\nu}/T_{\gamma}$ with respect to $T_{\gamma}$. Considering values of $g_{\mu}$ which are consistent with $\Delta a_{\mu}$, the decay width to photons leads to a decay rate higher than the Hubble rate at $T ={\cal{O}}$(1~MeV), heating the photon plasma. This is compensated for by lowering  $T_D$ when ${\rm BR}_{\nu\nu}\neq 0$ driving  
$T_\nu / T_\gamma$ back to values consistent with ${\rm \Lambda CDM}$~\cite{Planck:2018vyg}, $N_{\rm eff}=2.9913 \pm 0.1690$. We see in Fig.~\ref{fig:rate-Tratio} that this can be achieved with a small branching ratio to neutrinos ${\rm BR}_{\nu\nu}\simeq\mathcal{O}(10^{-2})-\mathcal{O}(10^{-3})$, corresponding to a coupling to neutrinos $g_{\nu_a} \simeq  {\rm few} \times 10^{-10}$ if we assume approximately equal couplings to all  neutrino flavors, for $m_\phi =1$--$2$~MeV. When ${\rm BR}_{\nu \nu} \ll 1$, the entropy injection of the scalar into the neutrino sector is negligible and its impact is implicit through the value of $T_D$.

Finally, we define $N_{\rm eff}$  at $T<T_D$  as the effective number of SM neutrinos that would lead to the same energy density contribution, including the electron and scalar $\phi$ contribution. However, for $T_\gamma \ll {\rm MeV}$, the contributions from $e$ and $\phi$ are exponentially suppressed and one can simply use Eq.~(\ref{eq:TnuTgamma}) in Eq.~(\ref{eq:neff}).
For ${\rm BR}_{\nu \nu} \ll 1$ and $T_\gamma \ll m_e,m_\phi$, we have 
\begin{equation}
N_{\rm eff} \approx N_\nu  \left(\frac{11/4}{1+ \frac{7}{4}F(\frac{m_e}{T_D})+\frac{7}{16}F(\frac{m_\phi}{T_D})}\right)^{4/3}.
\end{equation}
We emphasize that, for a given scalar mass, $T_D$ is determined by Eq.~(\ref{eq:TD_sol}). In Fig.~\ref{fig:TD2}, we show the range of acceptable values of $T_D$ and $m_{\phi}$ varying $\mathrm{BR}_{\nu\nu}$ to achieve $N_{\rm eff}$ within 2-$\sigma$ of the measured value. 
We see that for a scalar in the cosmological triangle, values of $T_D$ between 0.4~MeV and 0.8~MeV are needed to be consistent with the Planck measurements. Note that this range is safely above the lower bound from BBN, $T_D \gtrsim 0.3$ MeV~\cite{Sabti:2019mhn}.
\\

\noindent {\textbf{UV Model}}-- In this section, we present a UV completion to the effective model. We start by introducing a SM scalar singlet $\phi$ and doublet $H'$ with equal charges under a parity $Z_2$ symmetry and  two singlet Weyl fermions ($N$, $N'$), only one of which is charged under $Z_2$. In addition, we assume that the right-handed muon field is also charged under $Z_2$. The particle content and charges relevant for our discussion are summarized in Table~\ref{tab:U1PQMUUV} and the Lagrangian reads,
\small
\begin{align}
& \mathcal{L}_{\rm UV}= y_\mu \bar L_{\mu} H^{\prime}  \mu_{R}+ y_{N,i}  \left(L_i \cdot H \right)  N + y_{N',j}  \left(  L_j \cdot H' \right) N' \nonumber  \\
 &+ \lambda_N N N'\phi +  \mu_\phi H'^{\dag} H \phi  +\sum_{f=N,N'} \frac{m_f }{2} f  f + H.c. .
 \label{eq:LagUV}
 \end{align}
\normalsize
\begin{table}[htb]
  	\begin{tabular}{|c||c|c|c|c|c|c|}
 	\hline 
 	Field  & $\mu_R$  & $\phi$  & $H$ & $H'$ & $N$ & $N'$ \\  \hline
 	\hline
 	$SU(2)_L$ &  $1$   &  $1$ & $2$ & $2$ & $1$  & $1$   \\
 	\hline
 	$U(1)_Y$  &  $-1$  & $0$  & $1/2$ &  $1/2$ & $0$ & $0$   \\
 	\hline 
 	$Z_2$        &  $-1$  & $-1$  & $+1$ & $-1$  & $+1$  & $-1$   \\
 	\hline
 \end{tabular}
 	\caption{Particle content for the UV completion of the effective model. Added to the SM are a singlet (doublet) scalar $\phi$ ($H'$) and two singlet fermions $N$ and $N'$. A $Z_2$ parity symmetry is added. $\mu_R$ is the only particle in the SM which carries a $Z_2$ charge.
 	}
 	\label{tab:U1PQMUUV}
\end{table}
The $Z_2$ symmetry may be softly broken by mass terms in the scalar potential, to allow proper electroweak symmetry breaking with the required vacuum expectation value (vev) of the new doublet $H'$. 

The effective coupling of $\phi$ to muons is given by
 \begin{equation}
 g_{\mu}  = y_\mu \frac{ v \mu_\phi}{\sqrt{2} m_{H^{\prime}}^2}.
 \end{equation}
Hence, taking for instance $y_\mu \simeq 10^{-1}$ and $\mu_\phi \simeq 10$~GeV, $g_\mu$ is in the same order of the SM Higgs muon coupling for $m_{H^{\prime}} \simeq 700$~GeV. In order to obtain the proper muon mass, the vev of $H'$ should be $v'/\sqrt{2} \simeq 1$~GeV.  Such a scalar, which 
couples mostly to muons and neutrinos will not be detectable at current colliders, but could easily be seen at 
a future high energy muon collider with center of mass energies larger than $m_{H^{\prime}}$. 

We assume that the Dirac mass term $\lambda_N v_\phi$, coupling the sterile neutrinos $N$ and $N'$, is much larger than
the Majorana mass terms $m_N$ and $m_{N'}$.  In this limit, the mass eigenvalues for the three light active neutrinos and two sterile neutrinos $N$ and $N'$ are
approximately given by
\begin{align}
             &\tilde{m}_{\nu_1} = 0, \\
		 & \tilde{m}_{\nu_{2,3}} = \frac{v v' }{\sqrt{2} \lambda_N v_\phi} \left(|\vv{y_N}| |\vv{y_{N'}}| \mp  
		 \vv{y_N} \cdot \vv{y_{N'}} \right)\sim 0.1 {\rm eV},  \label{eq:neumass23}\\
	 & \tilde{m}_{N, N'} \approx \frac{\lambda_N v_\phi}{\sqrt{2}} \pm \frac{m_N+m_{N'}}{2}+
	 \mathcal{O}\left( \frac{1}{\lambda_N v_\phi}\right),
\end{align}
where $\vv{y_N}$ and $\vv{y_{N'}}$ are the vectors of Yukawa couplings in flavor space for the three generations of active neutrinos. 
One massless active neutrino exists because the mass matrix is rank four.
In a block-diagonal intermediate basis (see Supplemental Materials), the active neutrino mass matrix is then given by
\begin{equation} 
{\cal{M}}_{ij} = \left( y_{N,i} y_{N',j} + y_{N,j} y_{N',i} \right) \frac{\sqrt{2} v v'}{2 \lambda_N v_\phi}.
\label{eq:Mij}
\end{equation}
After diagonalizing this matrix, one obtains the eigenvalue masses of neutrinos and the mixing angles, which follow the usual seesaw expressions. 
Since the active neutrino mass is proportional to $1/v_\phi$, the $\phi$-$\nu$ coupling is, 
apart from an irrelevant sign,
\begin{align}
g_{\nu_a}&= \frac{\tilde{m}_{\nu_a}}{v_\phi}.
\end{align}
In our setup the neutrino spectrum must be 
hierarchical, with $m_{\nu_1} \simeq 0$, $m_{\nu_2} \simeq 8.7 \times 10^{-3}$~eV 
and $m_{\nu_3} \simeq 0.059$~eV, to be consistent with the current 
neutrino oscillation data and the present neutrino mass
bounds from laboratory data and cosmology \cite{Planck:2018vyg, ParticleDataGroup:2020ssz}.
The scalar will dominantly decay into the heaviest neutrino, $\nu_3$, with a
coupling  
\begin{equation}
g_{\nu_3} \simeq 2.5 \times 10^{-10},
\end{equation}
to be consistent with values of ${\rm BR}_{\nu\nu}$ needed for $N_{\rm eff}$, leading to $v_\phi \simeq 240$ MeV. 
Thus, in Eq.~(\ref{eq:Mij})
\begin{align}
\frac{y_{N,i}y_{N',j}v'}{ \lambda_N }\simeq  10^{-13}{\rm GeV},
\end{align}
impling that $y_{N,i}, y_{N',i}$ must be small.
Taking $\lambda_N = {\cal{O}}(1)$ allows us to identify the neutrino masses with $v_\phi$. In such a case, the sterile neutrinos
are significantly heavier than 1~MeV. Further, if in addition $y_{N,i}, y_{N',i}$ are of order of a few $10^{-7}$ 
the mixing between the sterile and active neutrinos will be considerable so that the sterile neutrinos will decay fast enough to avoid 
constraints from BBN. 
Finally, the scalar $\phi$ can couple to electrons via mixing with the SM Higgs, but the coupling can be easily suppressed down to
values of order $\sim 10^{-10}$, thus not affecting this scenario. 
\\

\noindent {\textbf{Conclusion}}--We have presented a model of a real scalar, with a mass in the MeV range, that
leads to the active neutrino mass generation and to an explanation of the observed muon anomalous
magnetic moment. Neutrino masses are generated through the addition of sterile neutrinos, with
masses in the hundreds of MeV range. 
This model is interesting phenomenologically, while also providing an example of the avoidance
of the cosmological constraints on MeV scalar particles coupled to photons. The relevant 
modification of the cosmological history is provided by the coupling to neutrinos, which should
be in a range leading to a branching ratio to neutrinos of order $(0.1-1)$\% in order to be consistent with observed number of relativistic degrees of freedom. 
It is worth mentioning that as the scalar we consider has a long lifetime its presence at colliders will appear in the form of missing energy. If the scalar potential contains a coupling to the SM Higgs boson through a quartic coupling this will lead to a contribution to the invisible decay of the Higgs, and therefore a complementary collider signal of the scenario we have presented.

\begin{acknowledgements}
\noindent {\textbf{Acknowledgments}}:
We thank David McKeen, Hongwan Liu and Julian Mu\~noz for useful discussions. C.W. would like to thank the Aspen Center for Physics, which is supported by National Science Foundation grant No. PHY-1607611, where part
of this work has been done. C.W. has been partially supported  by the U.S. Department of Energy under contracts No. DEAC02-06CH11357 at
Argonne National Laboratory.  The work of C.W. at the University of Chicago has been also supported by the DOE grant DE-SC0013642. TRIUMF receives federal funding via a contribution agreement with the National Research Council of Canada.
The work of J.L. is supported by National Science Foundation of China under Grant No. 12075005
and by Peking University under startup Grant No. 7101502597.
The work of X.P.W. is supported by National Science Foundation of China under Grant No. 12005009.
\end{acknowledgements}

\bibliography{ref}

\clearpage

\onecolumngrid

\fontsize{12pt}{14pt}\selectfont
\setlength{\parindent}{15pt}
\setlength{\parskip}{1em}

\begin{center}
	\textbf{\large A $\nu$ scalar in the early universe and $(g-2)_{\mu}$ } \\ 
	\vspace{0.05in}
	{ \it  Supplemental Material}\\ 
	\vspace{0.05in}
	{Jia Liu$^{1,2}$, Navin McGinnis$^{3}$, Carlos E.M. Wagner$^{4,5,6}$, Xiao-Ping Wang$^{7,8}$}
	\vspace{0.05in}
\end{center}
\small
\centerline{{\it  $^{1}$School of Physics and State Key Laboratory of Nuclear Physics and Technology, Peking University, Beijing 100871, China}}
\centerline{{\it  $^{2}$Center for High Energy Physics, Peking University, Beijing 100871, China}}
\centerline{{\it  $^{3}$TRIUMF, 4004 Westbrook Mall, Vancouver, BC, Canada, V6T 2A3}}
\centerline{{\it  $^{4}$High Energy Physics Division, Argonne National Laboratory, Argonne, IL 60439, USA}}
\centerline{{\it  $^{5}$\mbox{Physics Department and Enrico Fermi Institute, University of Chicago, Chicago, IL 60637}}}
\centerline{{\it  $^{6}$\mbox{Kavli Institute for Cosmological Physics, University of Chicago, Chicago, IL 60637, USA}}}
\centerline{{\it  $^{7}$School of Physics, Beihang University, Beijing 100083, China}}
\centerline{{\it  $^{8}$Beijing Key Laboratory of Advanced Nuclear Materials and Physics, Beihang University, Beijing 100191, China}}
\normalsize
\vspace{0.05in}
{In this supplemental material, we show the diagonalization procedure for the active and sterile neutrino mass matrix.}
\vspace{0.15in}

In the UV model, we have given the Yukawa coupling for the three active neutrinos and sterile neutrino $N, ~N'$.
To obtain the mass eigenstates and mixing matrices, one needs to  diagonalize the $5\times 5$ mass matrix. 
Since the mass matrix is rank four, one massless active neutrino is guaranteed.
The other eigenvalues have been given in the main text with the choice that the Dirac mass term $\lambda_N v_\phi \gg m_N, ~m_{N'}$. 
Therefore, the active neutrino masses $\tilde{m}_{\nu_i} $ are approximately independent of  $m_N$ and $m_{N'}$, because the large mass in the seesaw mechanism is actually $\lambda_N v_\phi$.
One can block diagonalize the neutrino mass matrix, separating the active neutrinos from the sterile neutrinos and simultaneously diagonalize the sterile neutrino $2\times2$ block matrix,
by performing the following rotation 
\begin{align}
	\tag{S1}
	\left( \begin{array}{c} \vv{\nu} \\ N \\ N'  \end{array}\right) \approx	
	\left(\begin{array}{ccc}
		\mathbb{I}_{\rm 3\times 3} &  \frac{\vv{y_N} v + \vv{y_{N'}} v'}{\sqrt{2} \lambda_N v_\phi} 
		&  \frac{\vv{y_N} v - \vv{y_{N'}}v'}{\sqrt{2} \lambda_N v_\phi} \\
		-\frac{ \vv{y_{N'}}^{\rm T} v'}{ \lambda_N v_\phi} &  \frac{1}{\sqrt{2}} + z &  -\frac{1}{\sqrt{2}} +z \\
		-\frac{\vv{y_N}^{\rm T} v }{ \lambda_N v_\phi} &  \frac{1}{\sqrt{2}} +z &  \frac{1}{\sqrt{2}} +z \\
	\end{array}\right)
	\left( \begin{array}{c} \vv{\nu}' \\ \tilde N \\ \tilde N'  \end{array}\right) .
	\label{eq:mixing}
\end{align}
Here $\vv{\nu }'$ are intermediate active neutrino states defined by the above rotation, while $\tilde{N}$ and $\tilde{N}'$ are already in their mass eigenstates. We have defined $z \equiv (m_N - m_{N'})/(4 \lambda_N v_\phi)$ and assumed $m_N > m_{N'}$ to derive the above rotation matrix.
The active neutrino mass matrix in the $\vv{\nu }'$ basis is given by
\begin{align} 
	\tag{S2}
	{\cal{M}}_{ij} = \left( y_{N,i} y_{N',j} + y_{N,j} y_{N',i} \right) \frac{\sqrt{2} v v'}{2 \lambda_N v_\phi},
\end{align}
which can be obtained directly by applying the rotation in Eq.~(\ref{eq:mixing}) or by integrating out the heavy sterile neutrinos. After diagonalizing this mass matrix, we can obtain the eigenvalue masses which shows the usual seesaw expression for the neutrino masses, associated with Majorana
masses of order $\lambda_N v_\phi$. The massless eigenstate is actually $\vv{v}'\cdot (\vv{y_N} \times \vv{y_{N'}})$.
From here one can obtain the proper mass eigenstates $\nu_a$, and the mixing angles for both active  and sterile neutrinos.

\end{document}